\documentclass{article}
\usepackage[T1]{fontenc}
\usepackage[utf8]{inputenc}
\usepackage{ismir} %
\usepackage{amsmath,cite,url}
\usepackage{graphicx}
\usepackage{color}

\usepackage{lipsum}  
\usepackage{graphicx}
\usepackage{amsmath}
\usepackage{array}
\usepackage[skins]{tcolorbox}
\usepackage{multirow}
\usepackage{arydshln}
\usepackage{csquotes}
\usepackage{dashrule}
\usepackage{booktabs}
\usepackage{tikz}
\usepackage{arydshln}
\usepackage{makecell}
\usepackage{siunitx}
\usepackage[bookmarks=false]{hyperref}

\usepackage{url}  %

\makeatletter
\g@addto@macro{\UrlBreaks}{\do\/\do\-}
\makeatother

\usetikzlibrary{shapes.geometric, arrows.meta, positioning, intersections, decorations.pathreplacing}

\usepackage{tabularray}
\definecolor{DeezerPurple}{HTML}{A238FF}
\definecolor{DeezerOrange}{HTML}{FF673D}
\definecolor{DeezerTeal}{HTML}{96F9F3}
\definecolor{DeezerAcid}{HTML}{FCFF60}
\definecolor{DeezerRed}{HTML}{FF0000}
\definecolor{DeezerTurbo}{HTML}{FFED00}
\definecolor{DeezerRose}{HTML}{FF0092}
\definecolor{DeezerLime}{HTML}{C2FF00}
\definecolor{DeezerRobinsEggBlue}{HTML}{00C7F2}
\definecolor{DeezerFrenchPass}{HTML}{C1F1FC}
\definecolor{DeezerTidal}{HTML}{EBFFAC}
\definecolor{DeezerCottonCandy}{HTML}{FFC2E5}
\definecolor{DeezerCornflowerLilac}{HTML}{FFAAAA}
\definecolor{DeezerWhite}{HTML}{FFFFFF}
\definecolor{DeezerSilver}{HTML}{BFBFBF}
\definecolor{DeezerGray}{HTML}{808080}
\definecolor{DeezerMineShaft}{HTML}{404040}
\definecolor{DeezerPalePurple}{HTML}{EAD1FF}
\definecolor{DeezerMauve}{HTML}{D09AFF}
\definecolor{DeezerElectricViolet}{HTML}{A238FF}
\definecolor{DeezerWindsor}{HTML}{6E14BD}
\definecolor{DeezerPalePink}{HTML}{FDDDFF}
\definecolor{DeezerWhitePointer}{HTML}{FBBBFF}
\definecolor{DeezerHeliotrope}{HTML}{F673FF}
\definecolor{DeezerSeance}{HTML}{C01FC3}
\definecolor{DeezerPaleRed}{HTML}{FFC3C3}
\definecolor{DeezerCoralRed}{HTML}{FF3D3D}
\definecolor{DeezerThunderbird}{HTML}{D71F14}
\definecolor{DeezerPaleOrange}{HTML}{FFDAC6}
\definecolor{DeezerFlesh}{HTML}{FFBB95}
\definecolor{DeezerOutrageousOrange}{HTML}{FF673D}
\definecolor{DeezerPunch}{HTML}{DB452C}
\definecolor{DeezerPaleAcid}{HTML}{FBFFB7}
\definecolor{DeezerPaleCanary}{HTML}{FDFF95}
\definecolor{DeezerLaserLemon}{HTML}{FCFF60}
\definecolor{DeezerCitron}{HTML}{90931E}
\definecolor{DeezerPaleTeal}{HTML}{ECFFFE}
\definecolor{DeezerPaleCanary1}{HTML}{C7FBF8}
\definecolor{DeezerFoam}{HTML}{96F9F3}
\definecolor{DeezerCalypso}{HTML}{2F7C90}
\definecolor{DeezerPaleBlue}{HTML}{E4E7FF}
\definecolor{DeezerPeriwinkle}{HTML}{AAB2FF}
\definecolor{DeezerBlue}{HTML}{3448FC}
\definecolor{DeezerCalypso1}{HTML}{2836B5}

\title{AI-Generated Song Detection via Lyrics Transcripts}

\multauthor
    {Markus Frohmann$^{1,2}$ \hspace{1cm} Elena V. Epure$^1$ \hspace{1cm} Gabriel Meseguer-Brocal$^1$}
    {{\bf Markus Schedl$^{2,3}$ \hspace{1cm} Romain Hennequin$^1$}\\
    $^1$ Deezer Research, Paris, France \quad $^2$ Johannes Kepler University Linz, Austria\\
    $^3$ Linz Institute of Technology, AI Lab, Austria\\
    {\tt\small research@deezer.com, \qquad \{markus.frohmann, markus.schedl\}@jku.at}
}

\def\authorname{M. Frohmann, E. Epure, G. Meseguer-Brocal, M. Schedl, R. Hennequin}

\begin{document}

\maketitle

\begin{abstract}
    The recent rise in capabilities of AI-based music generation tools has created an upheaval in the music industry, necessitating the creation of accurate methods to detect such AI-generated content.
    This can be done using audio-based detectors; however, it has been shown that they struggle to generalize to unseen generators or when the audio is perturbed.
    Furthermore, recent work used accurate and cleanly formatted lyrics sourced from a lyrics provider database to detect AI-generated music.
    However, in practice, such perfect lyrics are not available (only the audio is); this leaves a substantial gap in applicability in real-life use cases.
    In this work, we instead propose solving this gap by transcribing songs using general automatic speech recognition (ASR) models.
    Once transcribed, lyrics are again available in a text representation, and established AI-generated text detection methods can be applied.
    We do this using several detectors.
    The results on diverse, multi-genre, and multi-lingual lyrics show generally strong detection performance across languages and genres, particularly for our best-performing model using Whisper large-v2 and LLM2Vec embeddings.
    In addition, we show that our method is more robust than state-of-the-art audio-based ones when the audio is perturbed in different ways and when evaluated on different music generators.\footnote{Our code 
    is available at \url{https://github.com/deezer/robust-AI-lyrics-detection}.}
   
\end{abstract}

\section{Introduction}
\label{sec:introduction}
\begin{figure*}[h]
\centering
\begin{tikzpicture}[
    node distance = 1.8cm,
    block/.style = {draw, rectangle, minimum width=2.5cm, minimum height=0.8cm, align=center, font=\sffamily\small, text=black, fill=gray!20, rounded corners=3pt, inner sep=6pt},
    decision/.style = {draw, rectangle, minimum width=1.8cm, minimum height=0.8cm, align=center, font=\sffamily\small, text=black, fill=gray!20, rounded corners=3pt, inner sep=6pt},
    arrow/.style = {-{Latex[length=2.5mm, width=3mm]}, line width=0.7pt},
    feature/.style = {draw, minimum width=1.5cm, minimum height=2cm, fill=DeezerTeal, align=center, font=\sffamily\small, rounded corners=3pt,},
    icon/.style = {minimum size=1cm} %
]

\node (song) at (0,0) {\includegraphics[width=2.3cm]{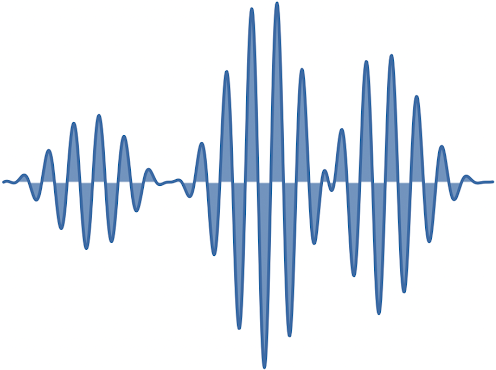}}; %
\node[above=0.1cm of song, font=\sffamily\bfseries\normalsize] (song-label) {Song};

\draw[arrow] (song.east) -- ++(2.3,0) node[midway, above=0.2cm, font=\sffamily\small] (transcriber-label) {Transcriber} coordinate (transcriber-arrow);
\node[below=0.4cm of transcriber-label] (transcriber) {\includegraphics[width=1.2cm]{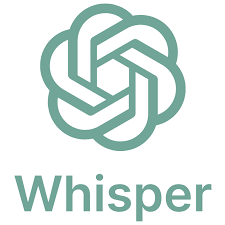}}; %

\node[block, right=0.1cm of transcriber-arrow] (transcription) {"Forever trusting\\ who we were and \\ nothing else shatters\\ Never cared for \\ what they shoe"\\...}; %
\node[above=0.3cm of transcription, font=\sffamily\bfseries\normalsize] (transcription-label) {Transcript};

\draw[arrow] (transcription.east) -- ++(2.3,0) node[midway, above=0.15cm, font=\sffamily\small, align=center] (feature-label) {Feature\\ Extraction} coordinate (feature-point);
\node[icon, below=0.2cm of feature-label] (feature-icon) {\includegraphics[width=1.6cm]{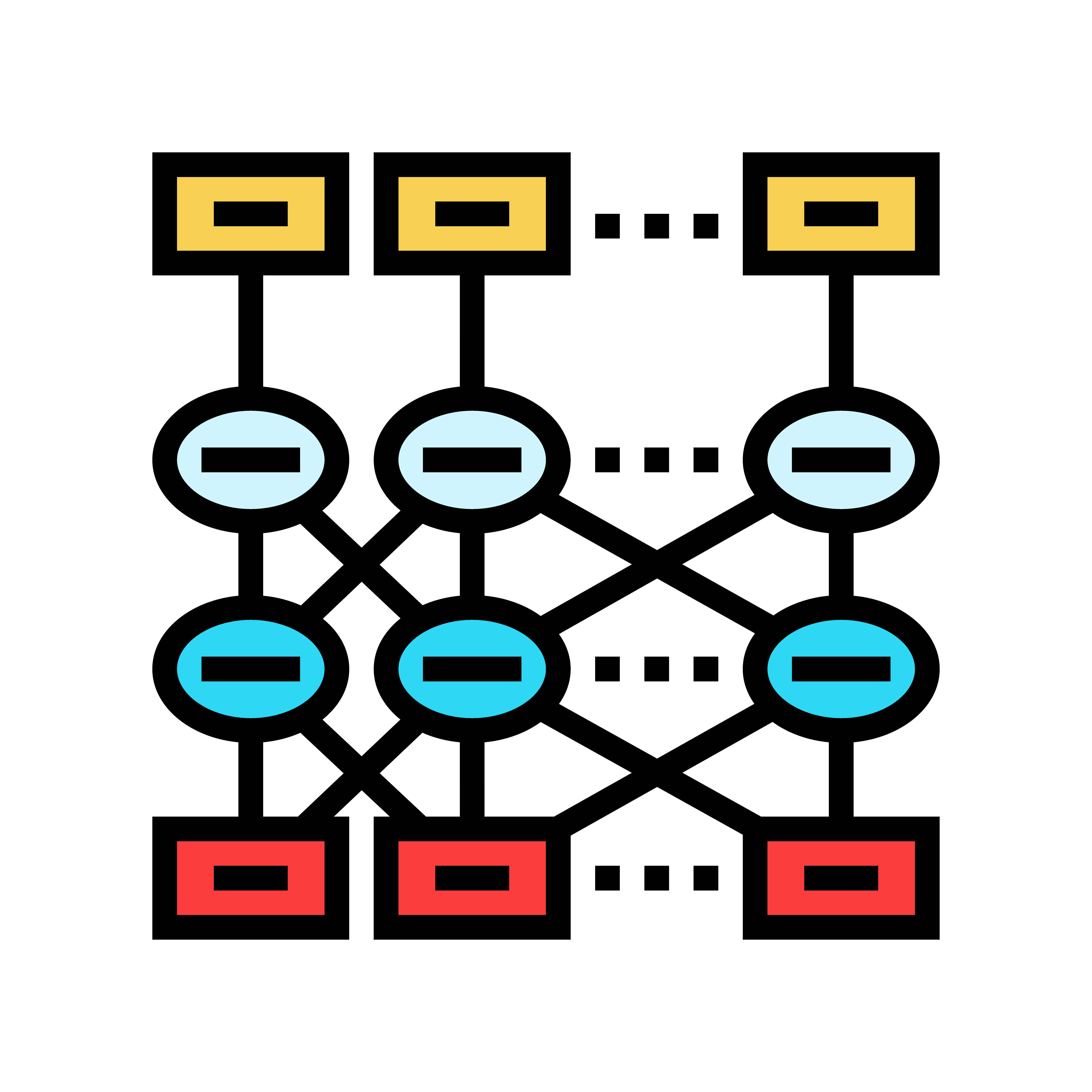}}; %

\node[feature, right=0.1cm of feature-point] (features) {
    0.12\\
    -0.30\\
    0.55\\
    0.08\\
    -0.15\\
    ...
};
\node[above=0.1cm of features, font=\sffamily\bfseries\normalsize] (features-label) {Features};

\draw[line width=0.7pt] (features.east) -- ++(1.5,0) node[midway, above, font=\sffamily\normalsize, xshift=0.5cm, yshift=0.3cm] (mlp-label) {MLP} coordinate (mlp-line);
\node[icon, below=0.4cm of mlp-label] (mlp-icon) {\includegraphics[width=1.4cm]{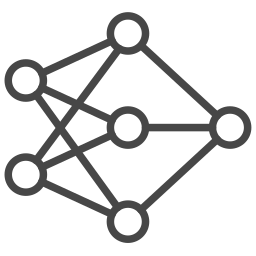}}; %

\draw[arrow] (mlp-line) -- ++(1.5, 0.75) node[decision, above right, fill=DeezerOrange] (fake) {\textbf{Fake}};
\draw[arrow] (mlp-line) -- ++(1.5, -0.75) node[decision, below right, fill=DeezerAcid] (real) {\textbf{Real}};
\vspace{-2mm}
\end{tikzpicture}
\vspace{-2mm}
\caption{Overview of our pipeline to detect AI-generated songs using lyrics. Using only its waveform, we get a song's lyrics using a \textit{transcriber} (e.g., Whisper). We then extract a feature vector from the transcript (e.g., with LLM2Vec), which is subsequently fed into an MLP-based \textit{detector} to classify the song as \textcolor{green}{real} or \textcolor{red}{fake}. Only the MLP classifier is trained while the other components are used as-is without further training.}
\vspace{-2mm}
\label{fig:pipeline}
\vspace{-2mm}
\end{figure*}
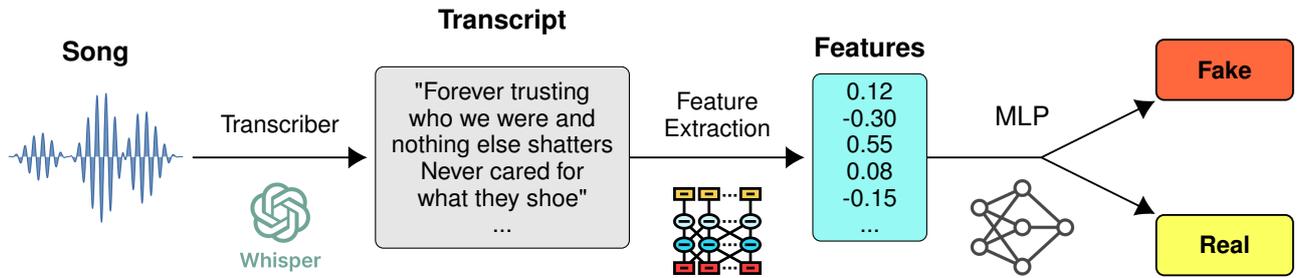

In recent years, the full generation of musical audio with artificial intelligence systems has matured \cite{Agostinelli2023MusicLMGM, copet2023simple,ning2025diffrhythm} and is now widely deployed in commercial systems such as \href{https://suno.com/}{Suno}, 
\href{https://www.udio.com/}{Udio}, 
\href{https://www.stableaudio.com/}{Stable Audio}, 
and \href{https://www.riffusion.com/}{Riffusion}.

The generation of this content presents new challenges for the music industry: 
Revenue dilution for real artists due to AI-generated tracks leading to profits \cite{genAI_german_chart}, copyright infringement issues \cite{RIAA_sues_suno_udio}, lack of transparency for the end user, or catalog flooding for music streaming services. For instance, Deezer reported that 10\% of the tracks delivered to them are generated by Udio or Suno, which is more than $10,000$ tracks daily \cite{deezer_genAI_MBW}.
Therefore, there is a pressing and growing need to identify this AI-generated content. 
While signed metadata \cite{c2paSpec} and watermarking \cite{ATSCAudioWaterMarkStandard} standards have been proposed to monitor and certify the source and history of content, they have not yet been widely adopted by the music industry. The only largely deployable solution currently remains automatic detection of this content.

Some recent methods \cite{afchar2024arxiv, afchar2025icassp} based on small CNN models have reported a very high detection accuracy (more than 99\%) for detecting AI-generated music audio files. However, these methods, primarily based on leveraging low-level artifacts of audio neural decoders, do not generalize to unseen generation models. In addition, they are very sensitive to audio manipulations such as playback speed modifications and can then be easily attacked.~\cite{afchar2024arxiv,afchar2025icassp}

Conversely, in \cite{labrak2024detectingsyntheticlyricsfewshot}, the authors propose a method for detecting AI-generated \textit{lyrics}, which shows promising results. 
Thus, lyrics derived from the audio signal could be leveraged to detect AI-generated content. 
As lyrics are mainly independent of the audio generation models, lyrics information should be mostly invariant under audio manipulation and audio generation model changes. Leveraging lyrics should lead to detection models that are more robust and generalize better.
However, the solution proposed in \cite{labrak2024detectingsyntheticlyricsfewshot} relies on the existence of properly formatted lyrics, which, in practice, are not available---only the audio is.\footnote{Lyrics are normally not provided as metadata when ingesting music in an industrial setting, which particularly affects new songs.}
This leaves a substantial gap in its applicability in real-life use cases.

In this paper, we confirm our hypothesis---that leveraging lyrics could lead to more robust and generalizable detection models---by proposing a new \textit{AI-generated music detection method that leverages lyrics} directly from audio via transcribing, as depicted in Figure \ref{fig:pipeline}. We show that this method maintains the state-of-the-art performance of \cite{labrak2024detectingsyntheticlyricsfewshot} across a diverse multi-genre corpus. Crucially, it does so despite the extra difficulty of transcription. Also, we show that it is \textit{robust to audio manipulation} and exhibits promising signs of \textit{generalization to unseen models}.

Finally, it should be noted that AI-generated audio and AI-generated lyrics are not perfectly correlated.
Many commercial models support inputting human-written lyrics; thus, it is possible to get AI-generated music with sung human-written lyrics. Moreover, it is possible to generate lyrics with AI and have them performed by real singers and orchestras.
Thus, the task of AI-generated lyrics detection and AI-generated music audio detection differs.
However, the second case---real music with AI-generated lyrics---is far less probable because investing significant resources in a professional recording with lyrics of limited perceived value is unlikely.
Overall, we believe detecting AI-generated lyrics can help with monitoring AI-generated content, as tracks with entirely AI-generated lyrics are likely to be fully AI-generated.
The task of AI-generated lyrics detection from audio would also have straightforward applications for publishers and copyright collecting societies to avoid providing royalties for AI-generated lyrics, as, for instance in the US, this content cannot be protected by copyright \cite{copyrightAct76}.

The paper is organized as follows:
In Section \ref{sec:related-work} we provide an overview of relevant domains: AI-generated music, and the detection of AI-generated music and text.
In Section \ref{sec:method}, we describe our method.
The experimental setup in Section \ref{sec:setup-experiments} focuses on the presentation of the data, baselines, and evaluation metric used in this work.
Then, we show results in Section \ref{sec:results} and conclude in Section \ref{sec:conclusion}.

\section{Related Work}
\label{sec:related-work}

\textbf{AI-generated music generation.}
Current AI music generation models typically rely on two main components working in sequence.
The first is an autoencoder (AE) trained to compress raw audio into a more manageable representation, which can then be reconstructed into an audio signal. Today, advanced neural audio codecs such as Encodec \cite{Defossez2022HighFNEncodec}, DAC \cite{kumar_23}, SoundStream \cite{Zeghidour2021SoundStreamAE}, Music2Latent \cite{marco_pasini_2024}, and MusicLM \cite{Agostinelli2023MusicLMGM}, are commonly used, enabling higher-quality generation. The second key component involves training a model to predict and generate the compressed sequence over time, often based on text prompts. Two principal approaches dominate this stage: large language models (LLMs), as explored in works such as \cite{ Zeghidour2021SoundStreamAE, Agostinelli2023MusicLMGM, copet2023simple}, and latent diffusion models, as seen in systems like Stable Audio \cite{Evans2024FastTL, Evans2024StableAO} and MusicLDM \cite{Chen2023MusicLDMEN}.  In simple terms, the AE is responsible for waveform synthesis, while the LLM or diffusion model ensures the generation of a coherent musical sequence over time. The commercial AI music-generation platforms, such as Suno, Udio, and Riffusion, generate entire songs, including lyrics conditioning. However, little is publicly known about the architectures of their underlying models.
\newline
\newline
\noindent
\textbf{AI-generated music detection.}
The first attempts to detect AI-generated content focused on detecting voice cloning \cite{zang2024singfake, desblancs2024real}. As these technologies continue to advance, it is increasingly difficult to distinguish cloned voices from real human performances. Consequently, researchers are focusing on identifying their presence in songs. A broader effort to detect AI-generated music has primarily focused on the AE component of music generators. Research such as \cite{afchar2024arxiv, afchar2025icassp} aims to determine if a music sample has been synthesized by an artificial decoder, independent of its musical content, by identifying artifacts introduced during the encoding-decoding process. 
More recently, the work of \cite{rahman2024sonics} has sought to identify whether the music and/or lyrics have been AI-generated and to determine which specific component was generated. 
However, audio-based methods have been shown to be prone to perturbations in audio, such as time stretching or adding noise \cite{afchar2024arxiv}, making their practical usage difficult.
\newline
\newline
\noindent
\textbf{AI-generated text detection.} Detecting whether a text is generated by AI has been widely researched. It is often framed as a supervised, binary classification challenge that consists of separating human-written content from machine-produced text~\cite{liu2023coco, huang2024toblend}. Most classifiers employ textual encoders such as RoBERTa or Longformer~\cite{abdelnabi2021adversarial, chakraborty2023counter, kirchenbauer2023watermark, liu2023coco, li-etal-2024-mage} or LLMs~\cite{macko2023multitude, antoun2024text, chen2023token, kumarage2023reliable}. However, this approach depends on having a sufficiently large training dataset, which is not always available, and it risks overfitting when faced with unfamiliar authorial styles or newly created generative models  \cite{uchendu2020authorship, bakhtin2019real}. A different strand of research aims to differentiate AI-generated text from human-written content by analyzing variations in generative model metrics or stylistic characteristics~\cite{Mitchell2023DetectGPTZM, su2023detectllm, zhu2023beat, sadasivan2023can}. Although these methods have shown effectiveness, their performance may vary compared to supervised approaches, influenced by the generative model and dataset employed ~\cite{li-etal-2024-mage}. Moreover, some researchers have investigated watermark-based detection techniques~\cite{abdelnabi2021adversarial, chakraborty2023counter, kirchenbauer2023watermark}. While these approaches yield promising results, most existing watermarking insertion schemes still require direct access to model logits, which external users of API-restricted models such as GPT-4 typically lack~\cite{achiam2023gpt}. Furthermore, \cite{dugan_etal_2024_raid} benchmark multiple AI-generated text detectors, showing that most of them exhibit high false positive rates, fail to generalize out-of-domain, and are vulnerable to different types of adversarial attacks, with each of the detectors showing different behavior.
While a plethora of research has explored AI-generated text detection across various domains, only \cite{labrak2024detectingsyntheticlyricsfewshot} have explored detecting AI-generated lyrics, introducing a corpus of synthetic lyrics generated using several LLMs with human lyrics seeds spanning nine languages from musically diverse genres.
In contrast, our aim is to detect AI-generated music via lyrics in a realistic scenario where \textit{only the audio} is accessible.

\begin{table}[t]
\begin{small}

    \centering
    \begin{tabular}{lccc}
    \toprule
      Text encoder  & Base model &  Dimension \\ \midrule
      \vspace{-2mm}\\
         \multicolumn{3}{c}{\textit{Retrieval-optimized conventional text encoders}} \\ 

         \textsc{MiniLMv2} & \textsc{XLM-Roberta} & 1024 \\
         \textsc{BGE-M3} & \textsc{XLM-Roberta} & 1024 \\
         \cdashline{1-3}

        \vspace{-2mm}\\
         \multicolumn{3}{c}{\textit{Stylistic text encoders}} \\
         \textsc{UAR-CRUD} & \textsc{distilroberta} & 768  \\

         \textsc{UAR-MUD} & \textsc{distilroberta} & 768  \\
        \cdashline{1-3}

        \vspace{-2mm}\\
        \multicolumn{3}{c}{\textit{LLM-based text encoders}} \\

        \textsc{BGE-ML-Gemma} &   \textsc{Gemma-2-9b} & 3584 \\
        \textsc{LLM2Vec-LLaMa} & \textsc{Llama-3-8b} & 4096 \\

         \bottomrule
    \end{tabular}
    \caption{Overview of the text encoders used.}
    \label{tab:text_encoders}
    \end{small}

\end{table}

\section{Method}
\label{sec:method}
\begin{table*}[t]
\begin{small}
\centering
\setlength\tabcolsep{8pt}
\begin{tabular}{l@{\hspace{12pt}} l S[table-format=3.0] S[table-format=3.0] S[table-format=3.0] S[table-format=3.0] S[table-format=3.0] S[table-format=3.0] S[table-format=3.0] S[table-format=3.0] S[table-format=3.0] S[table-format=3.0]}
\toprule
& & \texttt{en} & \texttt{de} & \texttt{tr} & \texttt{fr} & \texttt{pt} & \texttt{es} & \texttt{it} & \texttt{ar} & \texttt{ja} \\
\midrule
\multirow{2}{*}{\textbf{Human-written}} & Train & 30 & 30 & 30 & 30 & 30 & 30 & 30 & 30 & 30 \\
& Test  & 450 & 392 & 339 & 450 & 450 & 450 & 211 & 354 & 338 \\
\midrule
\multirow{2}{*}{\textbf{AI-generated}} & Train & 28 & 30 & 22 & 30 & 29 & 30 & 27 & 30 & 29 \\
& Test  & 450 & 450 & 241 & 450 & 450 & 450 & 150 & 407 & 255 \\
\bottomrule
\end{tabular}
\vspace{-1mm}
\caption{Distribution of human-written and AI-generated lyrics by language used as the generation seed (ISO 639 codes).}
\label{tab:stats-data}
\end{small}
\vspace{-1mm}
\end{table*}

\begin{table*}[h]
\begin{small}
\centering
\begin{tabular}{ccccccc}
\toprule
\texttt{fr}   & alternative & french & hip-hop   & pop                    & r\&b          & rock       \\
\texttt{it}   & alternative & electronic    & hip-hop   & jazz                   & pop           & rock       \\
\texttt{es}   & alternative & electronic    & hip-hop   & latin-american & pop           & rock       \\
\texttt{tr}  & alternative & electronic    & folk       & hip-hop               & pop           & rock       \\
\texttt{en}   & alternative & electronic    & hip-hop   & pop                    & r\&b          & rock       \\
\texttt{de}   & alternative & edm           & electronic & hip-hop               & pop           & rock       \\
\texttt{pt}   & christian   & hip-hop      & música popular brasileira        & pop                    & samba-pagode & sertanejo  \\
\texttt{ja}   & alternative & asian  & electronic & pop                    & rock          & soundtrack \\
\texttt{ar}   & alternative & arabic & electronic & hip-hop               & pop           & rock       \\ \bottomrule
\end{tabular}%
\vspace{-1mm}
\caption{Genres used for each language (ISO 639 codes). Selected genres are the most often streamed ones per language.}
\label{tab:list-genres}
\vspace{-2mm}
\end{small}
\end{table*}

The proposed pipeline to identify directly from audio if lyrics are AI-generated or human-written is illustrated in Figure \ref{fig:pipeline}. 
First, the audio is processed by a transcription model to generate a lyrics transcript. 
Building upon previous research showing their effectiveness with lyrics \cite{Zhuo2023LyricWhizRM,cifka2024lyrics} and robust to audio modifications\cite{katkov2024benchmarking}, we use pre-trained Whisper models, specifically \texttt{Whisper-large-v2} \cite{Radford2022RobustSR} using the \texttt{faster-whisper} library~\cite{guillaume_2023_fasterwhisper}.
The transcriptions are used as-is without any correction. 
We also experimented with various types of post-processing, such as text normalization, removing special characters, and stripping punctuation, but none of these improved detection performance.
Moreover, as shown in Section \ref{sec:results}, the performance gap between the best detector on ground-truth clean lyrics and transcribed noisy lyrics is small (less than 4\%). 
This suggests that even raw transcripts are promising for realistic scenarios where only audio is available.

We then input the complete lyrics transcript into a pre-trained multilingual text embedding model to capture semantic, syntactic, and stylistic properties. 
For a fair comparison, the context window is set to 512 tokens for all models. Most lyrics fit within this limit, but in rare cases where the token count exceeds 512, we truncate the input.
This step yields a single, contextualized vector-based representation of the lyrics.
Following past research on synthetic lyrics detection \cite{labrak2024detectingsyntheticlyricsfewshot}, we test multiple types of text embedding models: 
(1) conventional text encoders optimized for retrieval \cite{reimers-gurevych-2019-sentence,gao-etal-2021-simcse,chen-etal-2024-m3};
(2) LLM-based encoders~\cite{behnamghader2024llm2veclargelanguagemodels,chen-etal-2024-m3}; and
(3) text encoders designed to capture stylistic characteristics of text~\cite{rivera-soto-etal-2021-learning}.
While these models could be further specialized for lyrics, we leave this as future work;
although domain adaptation appears to help with the detection task \cite{labrak2024detectingsyntheticlyricsfewshot}, the overall performance gains remain modest, making it a lower priority for our work.
We provide a summary of the different models in Table \ref{tab:text_encoders}. %
\newline
\newline
\noindent
\textbf{Retrieval-optimized conventional text encoders}. The first category, \textit{retrieval-optimized conventional text encoders}, builds upon foundation models such as BERT \cite{devlin-etal-2019-bert} and MPNet \cite{song2020mpnet} and address some of their key limitations: high computational costs for tasks requiring the capturing of semantic similarity between pairs of textual inputs \cite{reimers-gurevych-2019-sentence}. 
In practice, the retrieval-optimized text encoder initializes a Siamese network with the weights of a foundation model such as BERT, which is then fine-tuned using a contrastive learning approach on pairs of similar texts.
The output is a fixed-size vector.

Although we tested multiple models from the \textsc{sentence-transformers} \cite{reimers-gurevych-2019-sentence} library, we report only the best-performing one for the detection task: a model based on \textsc{MiniLM} \cite{wang2020minilm}, a distilled version of \textsc{XLM-RoBERTa} \cite{conneau-etal-2020-unsupervised}, fine-tuned on over one billion similar sentence pairs.\footnote{For more details, see \url{https://www.sbert.net}.}
The second encoder, \textsc{BGE-M3} \cite{chen-etal-2024-m3} is built on the same foundation model, \textsc{XLM-RoBERTa}.
BGE-M3 introduces several innovations with regard to data curation---for both the self-supervised pre-training and supervised fine-tuning for sentence similarity and the training strategy---which relies on a self-knowledge distillation framework where multiple retrieval functions (embedding-based, aka dense retrieval; keyword-based, aka sparse retrieval) are learned together. 
\newline
\newline
\noindent
\textbf{LLM-based text encoders}.
The second category entails \textit{LLM-based text encoders}. LLMs are primarily designed for text generation, not text encoding, as they are decoder-based and trained autoregressively. 
However, recent work, such as LLM2Vec~\cite{behnamghader2024llm2veclargelanguagemodels}, has proposed using LLMs for text encoding, specifically via a three-step method to convert these models into text encoders.
First, the causal attention mask is modified to allow bidirectional attention.
Then, the model is pre-trained to adapt to this new attention mechanism with a Masked Next-Token Prediction (MNTP) objective.
Optionally, the model is further trained with contrastive learning in an unsupervised way using SimCSE~\cite{gao-etal-2021-simcse} to enhance sequence representation for downstream tasks. Specifically, the same input is subjected to multiple dropout masks to generate a pair of textual inputs used for sentence similarity fine-tuning.
This approach has shown strong results, making autoregressive LLMs very effective for text encoding but with a higher computational cost and inference time.

In our experiments, we used LLM2Vec based on \textsc{Llama-3-8b} and tested both the \textsc{MNTP} and \textsc{SimCSE} versions. 
Since their performance was similar in our task---consistent with the findings in \cite{labrak2024detectingsyntheticlyricsfewshot} on clean lyrics---we report results only for the \textsc{MNTP} model.
Additionally, we included another LLM-based model in our experiments, similar to \textsc{BGE-M3}, but built on Gemma \cite{team2024gemma} and optimized for multilingual text similarity and retrieval tasks: \textsc{BGE-ML-Gemma} \cite{chen-etal-2024-m3}\footnote{As of the time of writing, the detailed adaptation of Gemma in \textsc{BGE-ML-Gemma} for semantic text similarity has not been fully disclosed}.
\newline
\newline
\noindent
\textbf{Stylistic text encoders}. The third category consists of \textit{stylistic text encoders}, which have recently proven highly effective in identifying whether a text is human-written or generated by both open-source and closed-source LLMs \cite{sotofew}. 
Universal Authorship Attribution (UAR) models are trained to capture the author's writing style, complementing the usual syntactic and semantic cues in embeddings~\cite{rivera-soto-etal-2021-learning}. In practice, a contrastive learning strategy is used to train a retrieval-optimized conventional text encoder to separate style from topic. 
A positive example consists of an input from the same author on a different topic, while a negative example is from another author but on the same topic. 
UAR exists in multiple variants, \textsc{UAR-MUD} and \textsc{UAR-CRUD}, trained on input from 1 or 5 million Reddit users, respectively.

Finally, on top of the already extracted features with frozen pre-trained models, we train a multi-layer perceptron (MLP) with two hidden layers of sizes $256$ and $128$, using the ReLU activation function. 
We optimize the model with AdamW \cite{Loshchilov2017DecoupledWD}, setting the learning rate to $1e-3$ and reducing it by a factor of $0.1$ if the training loss does not improve for $5$ consecutive epochs. 
For all implementations, we use \texttt{pytorch-lightning} \cite{Falcon_PyTorch_Lightning_2019}.

\section{Experimental Setup}
\label{sec:setup-experiments}
We present further how we created the dataset starting from the lyrics-only corpus proposed by \cite{labrak2024detectingsyntheticlyricsfewshot}.
Then, we briefly describe the baselines and the evaluation metrics used.

\begin{table*}[]{
\small
\centering
\setlength\tabcolsep{10.5pt}
\centering
\begin{tabular}{lcccccccccc}

\textbf{Model} &  \texttt{en} & \texttt{de} & \texttt{tr} & \texttt{fr} & \texttt{pt} & \texttt{es} & \texttt{it} & \texttt{ar} & \texttt{ja} & \thead{\textbf{Macro}\\\textbf{Avg.}} \\

\toprule
\multicolumn{11}{c}{\textsc{Baselines}} \\
\midrule
$\textsc{GT Lyrics}_{\text{LLM2Vec}}$\textsuperscript{\textbf{\textdagger}}
 & 91.3 & 97.4 & 95.3 & \textbf{99.4} & 97.5 & 95.7 & 94.3 & 91.5 & 85.9 & 94.3 \\
$\textsc{CNN}_{\text{Spectrogram}}$\textsuperscript{\textbf{\textdaggerdbl}} & \textbf{97.5} & \textbf{96.3} & \textbf{97.5} & 98.7 & \textbf{98.8} & \textbf{97.0} & \textbf{98.0} & \textbf{94.4} & \textbf{98.4} & \textbf{97.4} \\
\midrule
\multicolumn{11}{c}{\textsc{Text-based detectors via lyrics transcriptions}} \\
\midrule

 \vspace{-5mm}\\

 \vspace{-2mm}\\
\multicolumn{11}{c}{\textit{Retrieval-optimized conventional text encoders}} \\
 \vspace{-2mm}\\

\textsc{MiniLMv2}  & 80.8 & 92.6 & 93.4 & 94.4 & 91.5 & 90.1 & 92.1 & 83.0 & 67.7 & 87.3 \\

\textsc{BGE-M3}  & 84.7 & 91.3 & 92.1 & 93.5 & 91.4 & 89.0 & 91.0 & 86.6 & 70.1 & 87.7 \\

\cdashline{1-11}

 \vspace{-2mm}\\
 \multicolumn{11}{c}{\textit{Stylistic text encoders}} \\
 \vspace{-2mm}\\

\textsc{UAR-CRUD} & 81.9 & 92.1 & 92.4 & 94.3 & 92.2 & 87.4 & 90.8 & 85.9 & 76.4 & 88.2 \\

\textsc{UAR-MUD}& 85.2 & 92.9 & 93.0 & 95.0 & 92.8 & 91.3 & 91.5 & 87.2 & 77.8 & 89.6 \\

\cdashline{1-11}

\vspace{-3mm}\\
 \vspace{-2mm}\\
\multicolumn{11}{c}{\textit{LLM-based text encoders}} \\
 \vspace{-2mm}\\

 \textsc{BGE-ML-Gemma} & 84.4 & 94.4 & 92.9 & \textbf{96.6} & \textbf{93.0} & \textbf{91.8} & \textbf{92.9} & \textbf{88.3} & \textbf{78.0} & 90.2 \\

\textsc{LLM2Vec-LLaMa} & \textbf{90.6} & \textbf{94.6} & \textbf{93.5} & 96.5 & 92.6 & 91.2 & \textbf{92.9} & 87.8 & 77.0 & \textbf{90.7} \\

\bottomrule

\end{tabular}%
}
\vspace{-2mm}
\caption{Recall scores for each language used as generation seed. We report the average over languages and the best lyrics-based model in \textbf{bold} per language.
\textsuperscript{\textbf{\textdagger}}~denotes the best-performing baseline by \cite{labrak2024detectingsyntheticlyricsfewshot}, using non-transcribed ground truth (GT) lyrics with $\textsc{LLM2Vec-LLaMa}$.
\textsuperscript{\textbf{\textdaggerdbl}}~uses the amplitude spectrogram to train a CNN on the task, following \cite{afchar2024arxiv}.}
\vspace{-2mm}
\label{tab:overall-results}
\end{table*}

\subsection{Data}
While several datasets with AI-generated audio exist, only the one by \cite{labrak2024detectingsyntheticlyricsfewshot} contains AI-generated \textit{lyrics}. 
It provides 3,704 real and 3,558 AI-generated lyrics
using three LLM generators (Mistral, TinyLlama, and WizardLM2) and human lyrics spanning nine languages and the six most popular genres per language as seeds.
Table \ref{tab:stats-data} presents the distribution of lyrics by language used as the generation seed, as well as by source (AI-generated or human-written) and train/test split.
The genres in Table \ref{tab:list-genres} correspond to the six most present music genres in each language, covering the majority of streams per language according to the statistics provided by the Deezer music streaming service. 

However, the dataset proposed by \cite{labrak2024detectingsyntheticlyricsfewshot} only provides lyrics; no audio is available. 
To enable realistic experiments representative of fully AI-generated music, we generate accompanying audio using Suno v3.5, conditioned on both lyrics and genre. 
It is capable of generating realistic songs with up to 4 minutes and represents the most recent stable Suno model.
Crucially, we utilize the provided lyrics from \cite{labrak2024detectingsyntheticlyricsfewshot} to ensure control over the content of the lyrics in the audio and to ensure diversity in terms of the generative model (LLM) used for the text modality.

For songs with human-generated lyrics, we use the original audio.
In experiments, we follow the same train/test split as introduced by \cite{labrak2024detectingsyntheticlyricsfewshot}.
Thus, our dataset contains diverse songs with (i) fake lyrics (LLM-generated) / fake audio (Suno-generated with lyrics conditioned from external LLMs) songs and (ii) real audio / real lyrics songs. 

Moreover, to assess generalization abilities to unseen audio and its artifacts, we want to test our already trained model on a new, previously unseen audio generation model. To this end, we turn toward Udio, another popular music generation tool, and generate 260 songs with AI-generated lyrics from the test set of the same lyrics dataset.
Here, we use the most recent \textit{udio-130 v1.5} model, setting $lyrics$ $strength$ to 100\% to ensure that the provided lyrics are used as-is without major changes and the seed to $42$. We leave the other settings at their default value. We then sample 260 real songs to maintain balance across languages and genres. We then evaluate the models trained on the Suno dataset on this out-of-domain scenario with both real and AI-generated songs.

\subsection{Baselines}
Our method considers a realistic scenario where only audio is available by leveraging transcribed lyrics to detect AI-generated music. We also include two strong baselines considering different scenarios.
The first, $\textsc{GT Lyrics}_{\text{LLM2Vec}}$, uses ground truth (i.e., non-transcribed; as found in a lyrics booklet) lyrics with the text encoding method \textsc{LLM2Vec} and Llama-3-8B as its underlying feature extractor since this combination performed best in \cite{labrak2024detectingsyntheticlyricsfewshot}. However, this assumes the availability of perfectly formatted lyrics. Like our models, an MLP is trained on the output embeddings, which are frozen. Second, we follow \cite{afchar2025icassp} and train a lightweight CNN on the audio's amplitude spectrogram directly on the task, aiming to detect audio artifacts.\footnote{Such models could also be trained on other input representations, but the findings of \cite{afchar2025icassp} are consistent across them. Hence, we resort to the best-performing one.}

\subsection{Evaluation}
We evaluate our model's performance using macro-recall, following \cite{labrak2024detectingsyntheticlyricsfewshot,nakov-etal-2013-semeval,li-etal-2024-mage}, as it provides a realistic evaluation across the two classes and is suitable for balanced datasets like ours.
Our focus is thus on minimizing false negatives (misclassifying real lyrics) and maximizing true positives (correctly identifying AI-generated lyrics).

\section{Results}
\label{sec:results}
We present the main experimental results in Table \ref{tab:overall-results}. 
As a reminder, the $\textsc{GT Lyrics}_{\text{LLM2Vec}}$ baseline refers to the method proposed by \cite{labrak2024detectingsyntheticlyricsfewshot}, which takes ground-truth lyrics as input and can be considered an upper baseline. 
The CNN method is a reimplementation of \cite{afchar2025icassp}, serving as a baseline that does not rely on lyrics information.

We observe only a slight performance decrease between the $\textsc{GT Lyrics}_{\text{LLM2Vec}}$ baseline and our proposed method using lyrics transcription when the text detection model is an LLM-based text encoder (BGE-Ml-Gemma and LLM2Vec-LLaMa).
This indicates that the lyrics transcription block effectively captures lyrical information, which is further exploited by our proposed method in Figure \ref{fig:pipeline} to classify whether a song is AI-generated or human-written when only audio is available.

Performance is consistent across all languages used as generation seeds, though we observe notable drops for Japanese and, to some extent, for Arabic and English as well. 
We analyze these errors and their sources in more detail in Section \ref{sec:qualitative_analysis}.
The other types of text encoders show more modest performance, confirming that the LLM2Vec encoder is a better feature extractor for AI-generated lyrics detection on both ground-truth and transcribed lyrics. 
The CNN baseline outperforms the lyrics-based model on in-domain data. 
However, its performance declines on out-of-domain data, as discussed in the next section (§\ref{sec:outofdomain}).

\subsection{Out-of-distribution Audio Generalization}
\label{sec:outofdomain}

We report the result of the out-of-distribution experiment in Table
\ref{tab:attack-results}: Under basic audio manipulations, the CNN method show a considerable performance drop for all transformations except time-stretching, similar to findings in \cite{afchar2025icassp}. 
In contrast, performance remains stable for lyrics-based detectors.
Moreover, lyrics-based models generalize well to unseen audio generators. 
Specifically, when trained only on Suno-generated audio, they still detect Udio-generated audio with only a small performance drop. 
In contrast, the artifact-based CNN model experiences a significant decline, performing only slightly better than chance on Udio-generated content.
This confirms the hypothesis that lyrics are largely unaffected by audio manipulation and can serve as a crucial extra cue for detecting fully generated content.

\subsection{Out-of-distribution Text Generalization}
Next, we assess generalization abilities of each model w.r.t. text generators.
For this, we keep music with lyrics generated by one of the models unseen from the training set and use it only at test time, while still using songs with lyrics from the other two models for training.
We report the results of this experiment in Table \ref{tab:generators-results}, where columns indicate the models that are not used during training but only at test.

First, the artifact-based CNN model maintains a strong performance when the lyrics generation model is changed ($95.3$\% on average). 
This was expected, as the lyrics generation model should not significantly affect the audio artifacts that the detection model relies on.
However, performance drops on music with lyrics generated by TinyLlama and WizardLM2, which is somewhat surprising. 

Considering lyrics-based models, their performance is more impacted by the set of lyrics generation models used in training. 
Yet, we still observe good generalization when the detection model is trained on data with lyrics from TinyLlama and WizardLM2 and tested on data with lyrics from Mistral ($88.9$\% for \textsc{LLM2Vec-LLaMa}). 
A larger performance drop occurs when testing on TinyLlama or WizardLM2 after training with the other two text generators.
Still, the performance remains decent and well above chance (unlike the audio-based CNN when tested on data from an unseen generator), indicating some generalizability to unseen generation models. 
Additionally, this suggests that training on diverse data from multiple text generation models (i.e., LLMs) is essential to maintain good performance on out-of-domain data.

\begin{table}[!t]{
\small
\centering
\setlength\tabcolsep{4.5pt}
\centering
\begin{tabular}{lccccc@{\hspace{2mm}\hspace{2mm}}c} %

\toprule
& \multicolumn{5}{c}{\textsc{Audio Attacks}} &  \\ 
\cmidrule(lr){2-6}
\multirow{2}{*}[1.5em]{\textbf{Model}} 
& \scalebox{0.83}{{Stretch}} & \scalebox{0.83}{{Pitch}} & \scalebox{0.83}{{EQ}} & \scalebox{0.83}{{Noise}} & \scalebox{0.83}{{Reverb}} 
& \multirow{2}{*}[1.5em]{\textsc{Udio}} \\
\midrule
\textsc{CNN} & \textbf{98.1} & 59.0 & 79.4 & 77.4 & 80.7 & 56.9 \\

\vspace{-3mm} \\
\cdashline{1-7}
\vspace{-2mm} \\

\textsc{UAR-MUD} & 86.7 & 88.8 & 88.8 & 88.6 & 88.5 & 85.6 \\
\textsc{BGE-Gemma} & 91.0 & \textbf{89.8} & \textbf{89.9} & \textbf{89.7} & \textbf{90.0} & \textbf{86.1} \\
\textsc{L2V-LLaMa%
} & 90.0 & 89.7 & 89.6 & 89.3 & 89.6 & 85.9 \\
\bottomrule

\end{tabular}%
}
\caption{Recall scores on out-of-distribution data (Udio) and when fake songs are perturbed (attacked) in different ways. We report average scores over languages.}
\vspace{-3mm}
\label{tab:attack-results}
\end{table}

\subsection{Qualitative Analysis}
\label{sec:qualitative_analysis}
In this section, we provide insights into the detectors' performance in identifying AI-generated lyrics when various pairs of genres and languages are used as seeds in generation. 
As a reminder, the most listened music genres per language are listed in Table \ref{tab:list-genres}.

As shown in Table \ref{tab:overall-results}, the best-performing model, \textsc{LLM2Vec-Llama}, effectively distinguishes AI-generated from human-written music overall. 
However, when examining the results in detail, performance varies across genre-language pairs.
In some cases, the model fails to detect a large portion (or even all) of the AI-generated content, particularly for Italian Jazz and Turkish Folk (both 0\% recall), as well as Japanese Electronic (32\%), Japanese Alternative (33\%), and Japanese Soundtrack (35\%) seeds.
In contrast, some English genres, such as Alternative, Electronic, Pop, R\&B, and Rock, exhibit higher false positive rates. 
This suggests that the model is more likely to misclassify music with human-written lyrics as AI-generated in these cases.

The low recall for certain genre-language pair seeds, such as Italian Jazz and Turkish Folk, can be largely attributed to severe class imbalance. Specifically, these pairs have very few AI-generated examples in the dataset
suggesting that insufficient training data is the determining factor in the model’s poor performance.
Yet, some English genres' high false positive rate may stem from stylistic similarities between human-written and AI-generated lyrics. 
A potential solution could be adjusting the detector’s classification threshold to be less aggressive in these cases, though we leave this for future work.

\begin{table}[!t]{
\small
\centering
\setlength\tabcolsep{5.2pt}
\centering
\begin{tabular}{lccccc}

\toprule
& \multicolumn{4}{c}{\textsc{Text generation models (LLMs)}} &  \\ 
& \multicolumn{4}{c}{\textsc{used during testing}} &  \\ 
\cmidrule(lr){2-5}

\textbf{Model} & \scalebox{0.9}{{Mistral}} & \scalebox{0.9}{{TinyLlama}} & \scalebox{0.9}{{WizardLM2}} & \thead{{\textsc{\textbf{Macro}}}\\{\textsc{\textbf{Avg.}}}} \\

\midrule
\textsc{CNN} & 99.4 & 92.6 & 94.0 & 95.3 \\

\cdashline{1-6}
\vspace{-2mm} \\
\textsc{UAR-MUD} & 85.2 & 71.3 & 78.1 & 78.2 \\
\textsc{BGE-Gemma} & 86.5 & 74.8 & 79.5 & 80.3 \\
\textsc{L2V-LLaMa%
} & 88.9 & 75.7 & 80.2 & 81.6 \\

\bottomrule

\end{tabular}%
}
\caption{
Recall scores of detectors (rows) when tested on lyrics from different generator LLMs (columns), following a leave-one-generator-out approach. Each model is trained on lyrics from the generators \textit{not} shown in the respective column. We macro-average scores over languages.
}
\vspace{-3mm}
\label{tab:generators-results}
\end{table}

\section{Conclusion}
\label{sec:conclusion}
In this work, we proposed a robust and practical method to detect AI-generated music focused on lyrics, but using only audio. 
To achieve this, we first transcribe the songs, overcoming the reliance on perfect ground-truth lyrics. 
Features are then extracted using various text encoders, and a lightweight MLP classifier is trained on top of these features to identify AI-generated music. 
Experiments show that the proposed method works effectively, is more resilient to audio distortions than past audio-only detection methods, maintaining performance where others degrade, and generalizes well to unseen AI music generation models, as well as fairly well to unseen LLMs for lyrics generation.
Future work should explore robustness to more complex audio manipulations and attacks, broader generalization to other AI music models once available, integrating alternative text encoders, including some fine-tuned on lyrics, and fusion with audio-based features for improved detection.
By enabling reliable detection directly from audio, our work offers a practical tool to address copyright concerns and ensure transparency. Its robustness and ease of deployment make it a practical solution for AI music management in the industry.

\section{Ethics statement}
Although designed for beneficial purposes such as safeguarding copyright and promoting transparency, the development and disclosure of AI detection systems pose complex ethical challenges. We acknowledge the dynamic landscape of this field and the need to catch up; as generative models evolve, their outputs may become statistically indistinguishable from human-created content. Detectors may be learning transient artifacts of current models, and their long-term utility is an open question.
Furthermore, our method addresses fully AI-generated content, but the growing prevalence of hybrid human-AI collaborative workflows presents a more nuanced scenario that current detectors do not yet address. In general, misapplication of such tools could lead to unwarranted removals of content, disproportionately harming creators. Moreover, biases embedded in training data may also skew detection outcomes across different musical genres or languages. To address these immediate and long-term challenges, we urge ethical development and implementation of AI-generated content detectors, prioritizing transparency about their capabilities and limitations, equitable outcomes, and human-in-the-loop approaches to balance innovation with protections for artists, creators, and the integrity of the music industry.

\section{Acknowledgements}
This research was funded in whole or in part by the Austrian Science Fund (FWF):  \url{https://doi.org/10.55776/COE12}, \url{https://doi.org/10.55776/DFH23}, \url{https://doi.org/10.55776/P36413}.
The authors would like to thank Aurelien Herault, Manuel Moussallam, Yanis Labrak, and Gaspard Michel for their invaluable feedback on this work.

\bibliography{neurips_2024,ACL2025}

\end{document}